\documentclass[10pt,twocolumn,letterpaper]{article}

\usepackage[pagenumbers]{cvpr} 

\usepackage{graphicx}
\usepackage{amsmath}
\usepackage{amssymb}
\usepackage{booktabs}

\usepackage{times}
\usepackage{epsfig}
\usepackage{bm}
\usepackage{float}
\usepackage{makeidx}  
\usepackage{verbatim}
\usepackage{bbding}
\usepackage{multirow}
\usepackage{pbox} 
\usepackage{cite}
\usepackage{threeparttable}
\usepackage[numbers,compress]{natbib}
\usepackage{xcolor}
\usepackage{caption}

\usepackage[pagebackref,breaklinks,colorlinks]{hyperref}

\usepackage[capitalize]{cleveref}
\crefname{section}{Sec.}{Secs.}
\Crefname{section}{Section}{Sections}
\Crefname{table}{Table}{Tables}
\crefname{table}{Tab.}{Tabs.}

\begin{document}

\title{One-shot Weakly-Supervised Segmentation in Medical Images}

\author{
Wenhui Lei$^{1,2}$, Qi Su$^1$, Ran Gu$^{2,3}$, Na Wang$^{4}$, Xinglong Liu$^{4}$, Guotai Wang$^{3}$, Xiaofan Zhang$^{1,2,*}$, \\ Shaoting Zhang$^{4}$ \\$^1$Shanghai Jiaotong University\\$^2$Shanghai AI Lab\\$^3$University of Electronic Science and Technology of China\\$^4$Sensetime Research \\ *Corresponding author: zhangxiaofan101@gmail.com
}

\maketitle

\begin{abstract}
Deep neural networks usually require accurate and a large number of annotations to achieve outstanding performance in medical image segmentation. One-shot segmentation and weakly-supervised learning are promising research directions that lower labeling effort by learning a new class from only one annotated image and utilizing coarse labels instead, respectively. Previous works usually fail to leverage the anatomical structure and suffer from class imbalance and low contrast problems. Hence, we present an innovative framework for 3D medical image segmentation with one-shot and weakly-supervised settings. Firstly a propagation-reconstruction network is proposed to project scribbles from annotated volume to unlabeled 3D images based on the assumption that anatomical patterns in different human bodies are similar. Then a dual-level feature denoising module is designed to refine the scribbles based on anatomical- and pixel-level features. After expanding the scribbles to pseudo masks, we could train a segmentation model for the new class with the noisy label training strategy. Experiments on one abdomen and one head-and-neck CT dataset show the proposed method obtains significant improvement over the state-of-the-art methods and performs robustly even under severe class imbalance and low contrast. 
\end{abstract}

\section{Introduction}
Precise automatic segmentation of medical images is crucial to various fields, e.g., surgical planning, radiation therapy, and other workflows \cite{lei2021automatic, wang2017automatic}. In recent years, deep learning-based segmentation algorithms have achieved superior performance with sufficient annotated data. However, collecting medical image segmentation annotations is expensive and time-consuming since it requires domain-specific experts to annotate at the pixel level. Furthermore, the data distribution shift between training and testing sets tends to hurt the generalisability of learned models, further aggravating the scarcity of labeled images \cite{castro2020causality}.
\begin{figure}[t!]
    \includegraphics[width=1\linewidth]{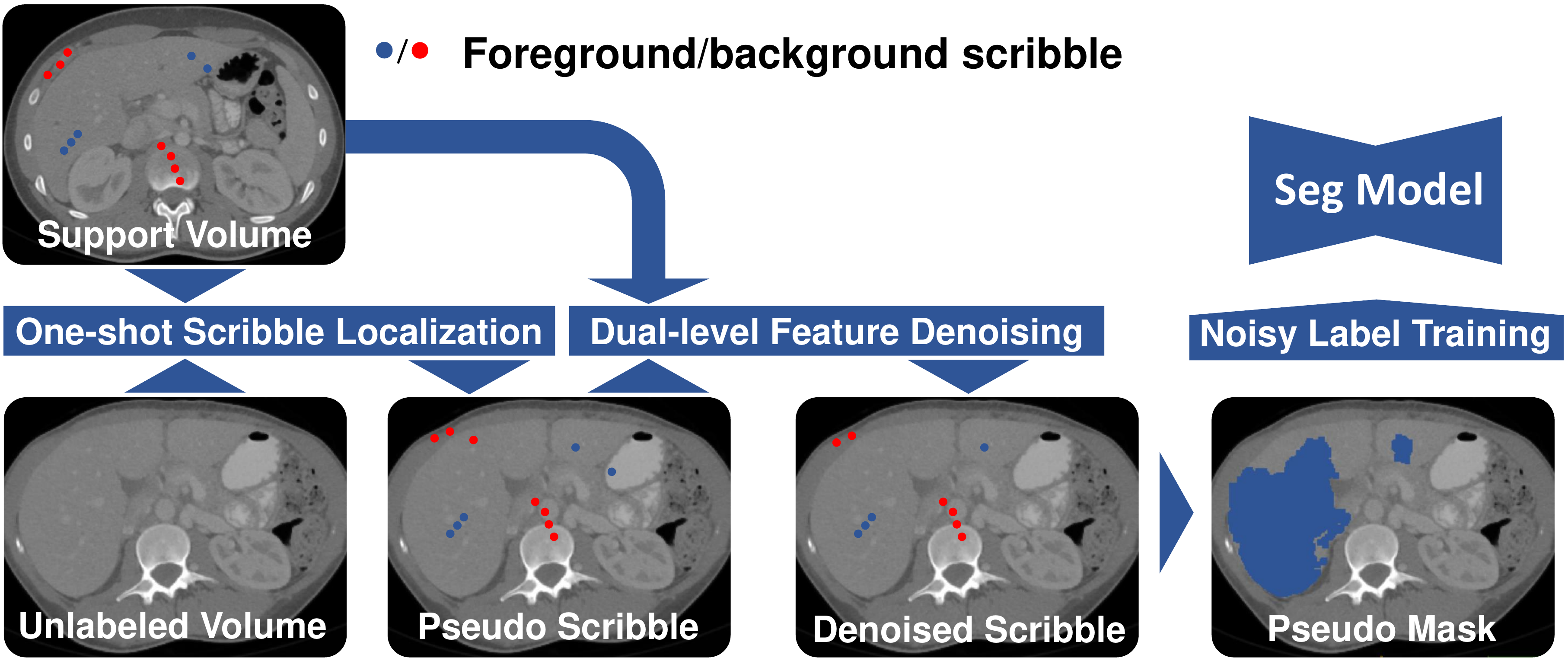}
   \caption{
   Our method tackles the one-shot medical image segmentation problem by localizing and refining scribbles in unlabeled volumes leveraging anatomical information and training the segmentation model with pseudo masks generated by scribbles using a noisy label training strategy.
   }
\label{fig:short}
\end{figure}
\par To overcome these challenges, plenty techniques have been investigated like semi-supervised- \cite{zhou2019collaborative}, self-supervised- \cite{jing2020self, lei2021contrastive, bai2019self}, weakly-supervised- \cite{hu2018weakly, dorent2021inter} and one-shot learning \cite{lei2021contrastive, ouyang2020self, zhao2019data}. Among them, one-shot learning is especially appealing because it only requires one annotated example (denoted as support) during the whole training stage and could segment the unlabeled images (denoted as query) in testing stage. Plenty of one-shot segmentation (OSS) methods have been proposed for natural image segmentation tasks \cite{wang2019panet, shaban2017one, caelles2017one, zhang2020sg}. They are mainly based on prototypical networks and mask average pooling to extract class prototypes from feature maps, with the assumption that that the prototypes contain enough information to distinguish the boundary around classes. However, different from natural images, medical images often suffer from (1) extreme sample imbalance between small foreground and large background area; (2) low contrast between foreground and surrounding tissues \cite{lei2021automatic}. Therefore, limited prototypes could hardly locate the target organs or separate the contour. Thus current medical image OSS methods based on prototypical networks \cite{roy2020squeeze, ouyang2020self} usually (1) require the range of target organs in one plane be given; (2) failed to segment the boundary in ambiguity area.

\par Therefore, we propose a novel OSS framework. We argue that previous works directly achieving dense segmentation results based on information from the support set may not be reasonable enough because they could not guarantee the accuracy of the contour voxels or even the correct localization of the target class. Since in most real clinical situations, there are much more unlabeled images than labeled ones \cite{tajbakhsh2020embracing}, it is more feasible to label limited points with high accuracy on unlabeled images, then expand them to formulate a large training set. 
\par More specifically, our method combines one-shot localization (OSL) with weakly-supervised segmentation (WSS), which only requires limited scribble annotations in the support image. First, we propose a propagation-reconstruction network (PRNet) to locate several foreground/background points in unlabeled images. Second, we design a dual-level feature denoising (DFD) method to select the correctly located points with anatomical- and pixel-level features. Then based on these points, we apply a WSS algorithm to achieve pseudo masks for the class-specific segmentation model training. Since the generated masks are not always accurate, we adopt a noisy training strategy to clean the label iteratively and eventually obtain a robust segmentation model.

\par Our contributions can be summarized as:
\par - A novel medical image OSS pipeline that combines OSL, WSS, and noisy training strategy to train segmentation models with only one weakly labeled image and several unlabeled images.
\par - PRNet for propagating scribbles to unlabeled images and DFD method for selecting the correctly located points.
\par - We demonstrate the effectiveness of our method on one abdomen CT dataset TCIA \cite{clark2013cancer} and one head-and-neck (HaN) dataset StructSeg19. Experiments show that the proposed framework outperforms the state-of-the-art few-shot framework for medical image segmentation largely by an average of 23\% on  TCIA and 45.4\% on StructSeg19 in terms of Dice score.

 \begin{figure*}[t!]
    \includegraphics[width=1\linewidth]{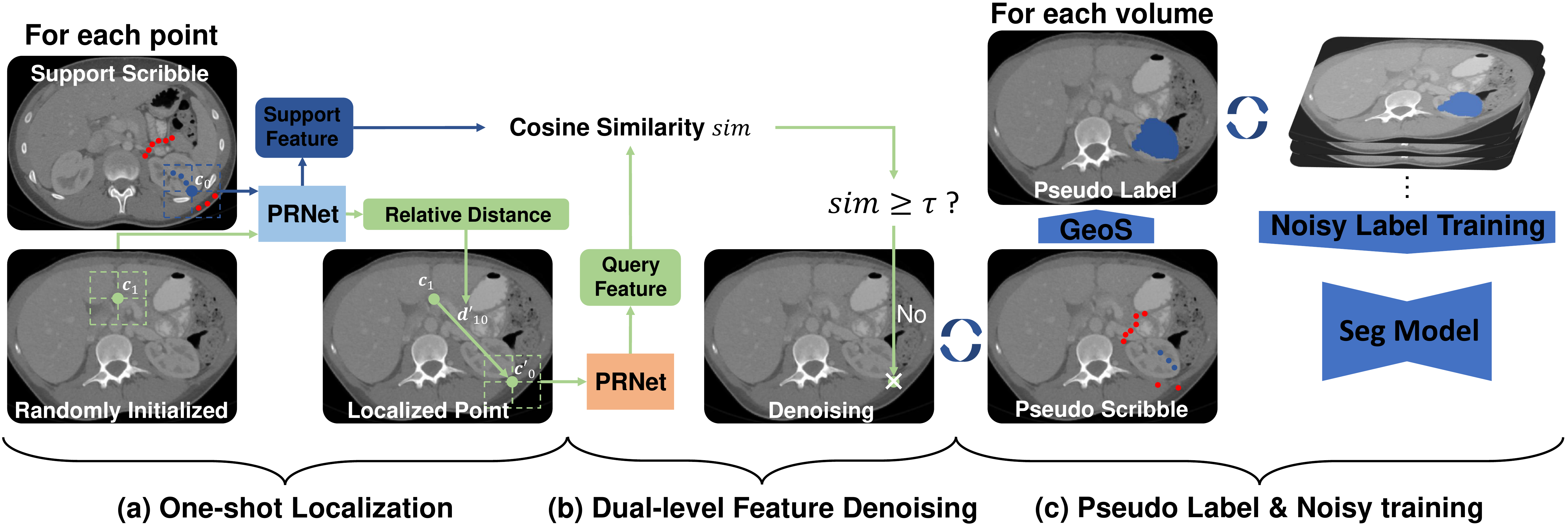}
   \caption{An overview of the proposed method. (a) One-shot scribble localization network for propagating the scribbles of support volume to the unlabele volume; (b) Dual-level feature denoising to refine the scribbles generated by OSL; (c) Generate pseudo masks with the scribbles GeoS method to train the segmentation task.}
\label{fig:train_frame2}
\end{figure*}
\section{Related Work}


\subsection{One-shot Learning:}
One-shot learning aims to identify a new category from only one training sample. Most recent works follow the research line of meta-learning, obtaining “generic” knowledge assumed to be shared among the known
and unseen classes \cite{finn2018probabilistic, finn2017model, snell2017prototypical}. 
These works can be roughly divided into three categories, i.e., the metric-based methods \cite{vinyals2016matching, sung2018learning, snell2017prototypical},  the model-based methods \cite{santoro2016one, munkhdalai2017meta}, and the optimization-based methods \cite{finn2017model, jamal2019task}.
\par OSS has achieved progressive success very recently \cite{shaban2017one, caelles2017one, rakelly2018few, wang2019panet, zhang2020sg, li2021adaptive}. A major stream of OSS network architecture in natural images is prototypical networks \cite{shaban2017one, caelles2017one}, which apply average mask pooling to generate one or multi-feature representations of fore-/background. Following the similar idea, Roy \etal~\cite{roy2020squeeze} first proposes a squeeze and excite architecture specifically for medical images few-shot segmentation. Tang \etal~\cite{tang2021recurrent} proposes a context relation encoder and a recurrent mask refinement module to refine the segmentation mask iteratively. However, they require the label of other organs around the target one for model training while collecting a large annotated training dataset for medical scans remains elusive due to the shortage of experts, reducing the feasibility of these methods. 

\par To leverage unlabeled data, self-supervised learning (SSL) attracted increasing interest because of its powerful ability in exploring the potential structure of medical image datasets \cite{bai2019self,lei2021contrastive, yao2021one, yan2020self, ouyang2020self, chaitanya2020contrastive}. Ouyang \etal~\cite{ouyang2020self} proposed a framework exploiting superpixel-based SSL, obtaining representation prototypes unsupervised and eliminating the need for manual annotations, while the model performance largely depends on the selection of superpixels.
\par Nonetheless, there are two shortages for the prototypical OSS methods mentioned above \cite{roy2020squeeze, ouyang2020self, tang2021recurrent}: 1) they focus on generating prototype representation of support images while neglecting the intrinsic information of the target class itself in query images, e.g., shape, size; 2) they are all 2D-based and need the start and end slice of target organ in one plane be given firstly in inference, which introduces additional supervision information.


Different from them, we expand our training set by localizing support scribbles on unlabeled images then obtaining pseudo masks through WSS algorithms to train the segmentation model. Recently, several works about medical images OSL \cite{lei2021contrastive, yao2021one, yan2020self} emerged, which proposed SSL methods for anatomical structure embedding. More specifically, \cite{yao2021one, yan2020self} compare the feature cosine similarity between target points and each pixel in query images for localization, while Relative Position Regression (RPR) \cite{lei2021contrastive} directly propagating patches to a shared physical 3D coordinate system, thus localizing the target point in a distance regression way. Although great progress has been made, current OSL methods do not have the self-checking mechanism and can not estimate the correctness of localization. Therefore, We extend the propagation networks (PNet) in RPR with an image reconstruction task to extract anatomical- and pixel-level features simultaneously for further filtering.

\subsection{Weakly Supervised Segmentation}
Compared with fully supervised segmentation task which needs time-consuming pixel-wise annotations, WSS requires more flexible annotations like scribbles \cite{lin2016scribblesup}, bounding boxes \cite{khoreva2017simple}, extreme points \cite{dorent2021inter} and image-level classification labels \cite{ahn2018learning}. Most previous WSS works in natural images achieve impressive performance by refining the CAMs \cite{zhou2016learning} generated by the image classifier to approximate the segmentation mask \cite{jo2021puzzle, ahn2018learning, huang2018weakly}. WSS also attracts great interest in the medical image analysis community due to its potential of reducing data labeling requirements, where the bounding boxes \cite{kervadec2019constrained, kervadec2020bounding, rajchl2016deepcut} and scribbles \cite{criminisi2008geos, wang2018deepigeos, lei2019deepigeos} are the most commonly used annotation type. Among all these methods, geodesic image segmentation (GeoS), which encodes spatial regularization and contrast sensitivity, has shown superior performance \cite{wang2018deepigeos, lei2019deepigeos, dorent2021inter}. Therefore, we adopt GeoS to form the pseudo masks used for supervision. However, generated masks may be inaccurate in practice, so we investigate noisy label learning to achieve better performance.

\par \subsection{Learning from Noisy Labels}
Many studies have shown that label noise can significantly impact the performance of deep learning models \cite{karimi2020deep, ren2018learning, berthelot2019mixmatch, han2018co, zhang2020weakly, wang2020noise, shi2021distilling}. Existing studies dealing with label noise could be organized under six categories: 1) Label cleaning and pre-processing \cite{ostyakov2018label, pham2019interpreting}; 2) Network architecture \cite{sukhbaatar2014learning, dgani2018training}; 3) Loss functions \cite{ghosh2017robust, matuszewski2018minimal}; 4) Data reweighting \cite{ren2018learning, le2019pancreatic}; 5) Data and label consistency \cite{lee2019robust, yu2019uncertainty}; 6) Training procedures \cite{zhong2019unequal, min2019two}. 
\par 
Karimi \etal~\cite{karimi2020deep} investigates the performance of different noisy-label learning approaches in approximate fetal brain segmentation generated by registration, intensity thresholding, and level set, which shares a similar situation with this propagate when training the segmentation model from pseudo masks. The experiment results show that the iterative label cleaning strategy achieves the best performance. Following this conclusion, we utilize a state-of-the-art cleaning training method \cite{zhang2021learning} that iteratively corrects labels with predicted probabilities above a decreasing threshold.

\section{Method}
In this section, we first introduce the problem setting and an overview of our framework. Then we elaborate details of it. To be simplified, we focus on binary segmentation, which could be easily extended to multi-classes.
\subsection{Problem Formulation}
Let $f_{\theta}$ be a CNN parametrized by the weights $\theta$ that predicts the probability of being foreground for each voxel. $\{\bm{X}_u\}$ represents a set of unlabeled grayscale 3D medical image scans, and $[\bm{X}_s, \bm{Y}_s]$ represents a support volume with its scribble annotation. We focus on the challenging setting of OSS that trains $f_{\theta}$ one support 3D image scans $[\bm{X}_s, \bm{Y}_s]$ and a set of unlabeled cases $\{\bm{X}_u\}$. 

As shown in Fig. \ref{fig:train_frame2}, our approach contains three parts:

\par 1) For each point $\bm{c}_0$ in the support scribble and a randomly selected point $\bm{c}_1$ in the unlabeled volume, PRNet takes the patches around them respectively to predict their relative distance $\bm{d}'_{10}$ and keep the support feature vectors. Then we move $\bm{c}_1$ to $\bm{c}_0'$ with $\bm{d}_{10}$ as a temporally located point.
\par 2) We crop the patch around $\bm{c}'_0$ to PRNet and get the query feature vectors. Then we calculate the cosine similarity $sim$ between support and query feature vectors. If $sim$ surpasses $\tau$, the $\bm{c}'_0$ would be kept, otherwise be deleted. We go through all the points in scribble and achieve the final propagated results.
\par 3) With the pseudo scribbles, we apply geodesic distance-based weakly supervised segmentation method \cite{wang2018deepigeos, criminisi2008geos} generating pseudo masks for each subject then train a segmentation model.

\subsection{Scribble Localization}
 \begin{figure*}[ht]
    \includegraphics[width=1\linewidth]{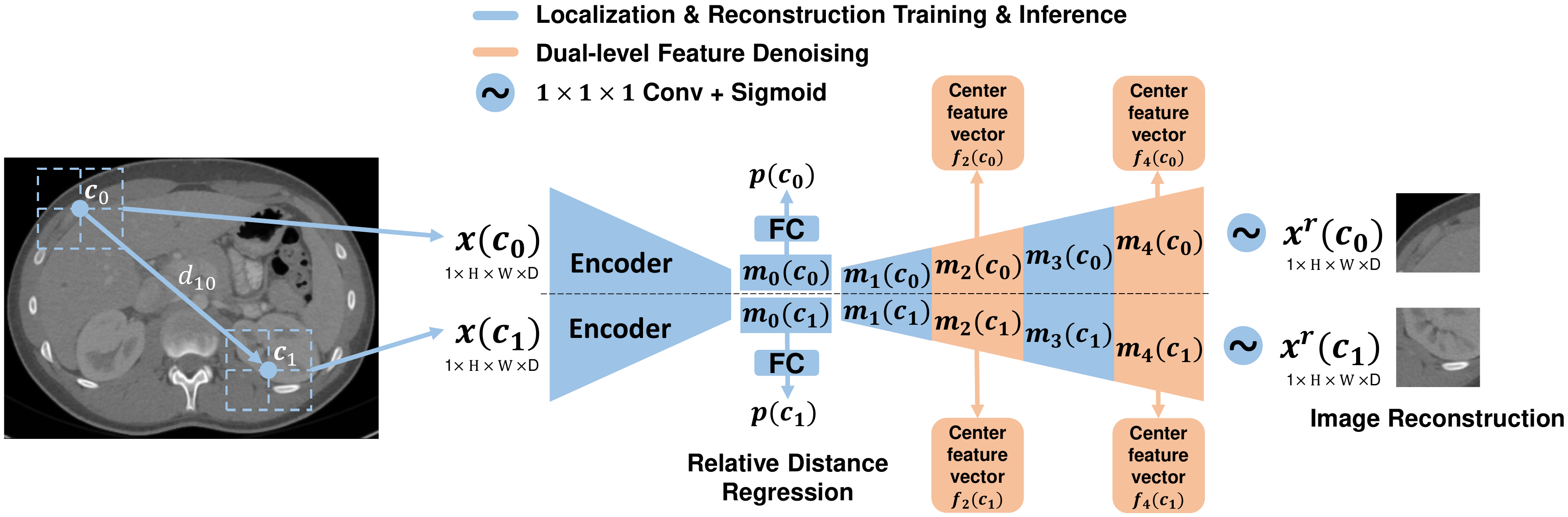}
   \caption{PRNet is composed of three parts: an encoder, fully connected layers and a decoder. The encoder contains four blocks of convolutions and $2\times2\times2$ downsampling while the decoder contains four blocks of convolutions and $2\times2\times2$ upsampling. $\bm{m}_i$ represents the feature map after $i$ times upsampling. We train PRNet with relative distance regression and image reconstruction: with two randomly selected points $\bm{c}_0$ and $\bm{c}_1$ from the same volume, we crop patches $\bm{x}(\bm{c}_0), \bm{x}(\bm{c}_1)$ and pass them into PRNet to get their anatomical coordinate predictions $\bm{p}(\bm{c}_0)$, $\bm{p}(\bm{c}_1)$ and reconstruction results $\bm{x}^r(\bm{c}_0)$, $\bm{x}^r(\bm{c}_1)$. Then we calculate the loss with Eq. \ref{eq3}. After training, we apply PRNet to propagate every point from support scribbles to query volumes. And to filter out correctly located points, we use DFD to the center corresponding feature vectors from $\bm{m}_2$ and $\bm{m}_4$ of the support point and the located point. }
\label{fig:PRNet}
\end{figure*}
Because a scribble could be viewed as a set of adjacent points, it is easy to obtain their counterparts in the query volume based on one-shot landmark localization frameworks \cite{lei2021contrastive, yao2021one, yan2020self}. However, due to the large variance among individuals, the located points may not be accurate enough and fall into the wrong areas, decreasing the segmentation performance. To resolve this issue and refine the noisy propagation, we propose a self-supervised feature similarity-based method.
\par RPR \cite{lei2021contrastive} propagates the medical scan patch into a shared 3D coordinate system, representing its anatomical position in the human body. We further assume that the information a medical scan patch holds could be disentangled as anatomical-level position (where the patch is) and the pixel-level representation (what kind of tissue the patch contains). Thus for a correctly located point, its features should be similar to the support point on both two levels, and the key falls in extracting the anatomical- and pixel-level feature from unlabeled data. 
\par Thus, we extend RPR with an image reconstruction task to extract pixel-level features simultaneously. More specifically, we add a decoder in the PNet of RPR for image reconstruction and design a propagation-reconstruction network (PRNet). The structure of PRNet is shown in Fig. \ref{fig:PRNet}, in which $\bm{m}_i$ means the feature map after $i$ times upsampling. We train the PRNet with two SSL tasks: relative distance regression and image reconstruction. 
\par During the training stage, we randomly select a volume $\bm{X}_u$ from unlabeled set and two points  $\bm{c}_0(\bm{c}_0^z,\bm{c}_0^x,\bm{c}_0^y), \bm{c}_1(\bm{c}_1^z,\bm{c}_1^x,\bm{c}_1^y)$ from it. Then we crop two patches $\bm{x}(\bm{c}_0)$, $\bm{x}(\bm{c}_1)$ around $\bm{c}_0$, $\bm{c}_1$ with fixed size $H\times W\times D$, respectively. Assuming the pixel spacing of $\bm{X}_u$ is $\bm{e}\in R^3$, the ground truth offset $\bm{d}_{10}$ from $\bm{x}(\bm{c}_0)$ to $\bm{x}(\bm{c}_1)$ in the physical space is denoted as: 
 \begin{equation}
    \bm{d}_{10}= (\bm{c}_1-\bm{c}_0) \circ \bm{e}
    \label{eq1}
\end{equation}
 where $\circ$ represents the element-wise product.
 \par We send $\bm{x}(\bm{c}_0)$, $\bm{x}(\bm{c}_1)$ to PRNet and obtain 4 items: (1) anatomical 3D coordinate predictions $\bm{p}(\bm{c}_0)$, $\bm{p}(\bm{c}_1)$; (2) image reconstruction $\bm{x}^r(\bm{c}_0)$, $\bm{x}^r(\bm{c}_1)$; (3) anatomical-level feature vectors $\bm{f}_2(\bm{c}_0)$, $\bm{f}_2(\bm{c}_1)$ from the center of feature maps $\bm{m}_2(\bm{c}_0)$, $\bm{m}_2(\bm{c}_1)$; (4) pixel-level feature vectors $\bm{f}_4(\bm{c}_0)$, $\bm{f}_4(\bm{c}_1)$ from the center of feature maps $\bm{m}_4(\bm{c}_0)$, $\bm{m}_4(\bm{c}_1)$. The last two items will be used for judging the correctness of located points in Sec. \ref{sec 3.3}. Then the predicted offset $\bm{d}'_{10}$ from $\bm{c}_1$ to $\bm{c}_0$ is obtained as:
 \begin{equation}
    \bm{d}'_{10}= r \cdot tanh(\bm{p}(\bm{c}_0)-\bm{p}(\bm{c}_1))
    \label{eq2}
\end{equation}
where the hyperbolic tangent function $tanh$ and the hyper-parameter $r$ together control the upper and lower bound of dqs, which is set to cover the largest possible offset. Finally, we apply the mean square error (MSE) to measure the relative distance and reconstruction loss:
\begin{align}
\begin{aligned}
    L_{ssl} = &L_{dis} + L_{rec} \\
    L_{dis}  = &\frac{1}{3}||\bm{d}_{10} - \bm{d'}_{10}||_2^2 \\
    L_{rec} = \frac{1}{N} (||\bm{x}(\bm{c}_0) -  &\bm{x}^r(\bm{c}_0)||_2^2+||\bm{x}(\bm{c}_1) -  \bm{x}^r(\bm{c}_1)||_2^2)
\end{aligned}
\label{eq3}
\end{align}
where $N = H\times W\times D$ is the number of total voxels of the patch.
\par After self-supervised training, the network can be directly used for localization on any landmark contained in the training dataset.
\par 
\label{sec 3.3}
Given a support volume $[\bm{X}_s$, $\bm{Y}_s]$, our mission is localizing every point in $\bm{Y}_s=\{y(\bm{c}_0), y(\bm{c}_1) ... , y(\bm{c}_i) ... \}$ on unlabeled volume set $\{\bm{X}_u\}$, in which $y(\bm{c}_i)$ represents the label of point $\bm{c}_i$. 
\par
We start from point $\bm{c}_0$ and traverse around the scribble. By the same token with training stage, we first crop patch $\bm{x}(\bm{c}_0)$ around $\bm{c}_0$ from  $\bm{X}_s$ and pass it through PRNet to achieve the corresponding anatomical coordinate $\bm{p}(\bm{c}_0)$, feature vectors $\bm{f}_2(\bm{c}_0)$ and $\bm{f}_4(\bm{c}_0)$. Then given an unlabeled image $\bm{X}_u$, we randomly crop a patch $\bm{x}(\bm{c}_1)$ with center point $\bm{c}_1$ to get $\bm{p}(\bm{c}_1)$, $\bm{f}_2(\bm{c}_1)$ and $\bm{f}_4(\bm{c}_1)$. The relative distance $\bm{d}'_{10}$ from $\bm{c}_1$ to $\bm{c}_0$ is obtained with Eq. (\ref{eq2}). 
 Thus the located point $\bm{c}'_{0}$ can be obtained by moving $\bm{c}_1$ with $\bm{d}'_{10}$:
  \begin{equation}
     \bm{c}'_{0}=\bm{c}_1+\bm{d}'_{10}
     \label{eq7}
 \end{equation}

 \subsection{Dual-level Feature Denoising} 

\par 
To validate the correctness of $\bm{c}'_{0}$, we propose dual-level feature denoising (DFD): we first crop patch $\bm{x}(\bm{c}'_{0})$ from $\bm{X}_u$ and feed it into PRNet, getting corresponding two level feature vectors $\bm{f}_2(\bm{c}'_{0})$ and $\bm{f}_4(\bm{c}'_{0})$. 
The theoretical reasoning and experiments in \cite{chen2019understanding} indicate a quadratic relation between the label noise ratio in the training data and test error. Consequently, we focus on improving the precision of located points rather than the total number. 
\par
A key observation of this research is that the feature vectors of located point and the support point are highly comparable in the lower level feature maps near the fully connected layers, 
i.e., $\bm{m}_0(\bm{c}_0), \bm{m}_0(\bm{c'}_0)$ and $\bm{m}_1(\bm{c}_0), \bm{m}_1(\bm{c'}_0)$. 
Because the predicted anatomical coordinate $\bm{p}(\bm{c}_0), \bm{p}(\bm{c}'_{0})$ are fully depended on the $0$ level feature map $\bm{m}_0(\bm{c}_0), \bm{m}_0(\bm{c'}_0)$. $\bm{p}(\bm{c}_0)\approx\bm{p}(\bm{c}'_{0})$ implies the equivalence of $\bm{m}_0(\bm{c}_0)$ and $\bm{m}_0(\bm{c'}_0)$. The low-level feature maps in the decoder hold anatomical information, and the high-level feature maps represent pixel-level information. 

Therefore, we select $\bm{f}_2(\bm{c}_0), \bm{f}_2(\bm{c'}_0)$ in $\bm{m}_2(\bm{c}_0), \bm{m}_2(\bm{c'}_0)$ and $\bm{f}_4(\bm{c}_0), \bm{f}_4(\bm{c'}_0)$ in $\bm{m}_4(\bm{c}_0), \bm{m}_4(\bm{c'}_0)$ from the support and query volumes, and calculate the cosine similarity of these two level features, then multiply them to get variable $sim$, quantifying the similarity of $\bm{c}_0$ and $\bm{c}'_{0}$ in anatomical and pixel-level simultaneously:
  \begin{equation}
     sim= cos(\bm{f}_2(\bm{c}_0), \bm{f}_2(\bm{c'}_0)) \cdot cos(\bm{f}_4(\bm{c}_0), \bm{f}_4(\bm{c'}_0))
     \label{eq7}
 \end{equation}
$\bm{c}'_{0}$ will be labeled as class $y(\bm{c}_0)$ if $sim>\tau$, otherwise be discarded. Eventually, we obtain the scribble propagation $\bm{Y}_u$ for each $\bm{X}_u$. 
In the inference stage, the same process could be applied to get the pseudo scribble $\bm{Y}_q$ from the given query volume $\bm{X}_q$.
 \begin{table*}[t!]
\centering
\caption{Dice score (\%) comparison in testing set.}
\label{tab:point_patch}
\scalebox{0.84}{
\begin{tabular}{@{}cccccccccc@{}}
\toprule[1pt]
& & \multicolumn{4}{c}{TCIA}& \multicolumn{4}{c}{StructSeg19} \\
Method & Manual Local.? & Spleen  & Left Kidney  & Liver & Mean & Brain Stem  & Left PG  & Right PG & Mean \\ \midrule[1pt]
SE-Net \cite{roy2020squeeze}  & \CheckmarkBold   & 32.9$\pm$10.6 & 30.5$\pm$16.2 & 54.0$\pm$6.9  & 39.1$\pm$9.1 & 4.6$\pm$1.8 & 2.2$\pm$0.5 & 2.0$\pm$0.5 & 2.9$\pm$0.7\\
SSL-ALPNet \cite{ouyang2020self}  & \CheckmarkBold & 57.5$\pm$12.0 & 63.9$\pm$10.6 & 75.6$\pm$4.4 & 65.7$\pm$6.4 & 23.9$\pm$4.0 & 13.6$\pm$4.4 & 20.0$\pm$6.5 & 19.1$\pm$3.9 \\
Aug \cite{zhao2019data}  & \XSolidBrush & 36.3$\pm$19.0 & 16.4$\pm$25.8 & 81.2$\pm$14.8 & 44.6$\pm$18.4 &55.6$\pm$10.6&30.2$\pm$13.2&33.8$\pm$15.7&39.9$\pm$12.4 \\
PRNet (ours) & \XSolidBrush & \textbf{ 84.9$\pm$12.0} & \textbf{90.9$\pm$4.1}  & \textbf{90.3$\pm$3.6} & \textbf{88.7$\pm$4.2} & \textbf{75.2$\pm$3.4} & \textbf{74.6$\pm$3.3} & \textbf{76.0$\pm$2.8} & \textbf{75.3$\pm$3.2} \\ \cmidrule(lr){1-10}
 Fully Supervised & \XSolidBrush & 90.2$\pm$14 & 95.0$\pm$1.6 & 93.9$\pm$4.8 & 93.0$\pm$6.3 & 82.9$\pm$3.6 & 85.8$\pm$2.1 & 84.9$\pm$4.0 & 84.9$\pm$3.5 \\ 
\bottomrule[1pt]
\end{tabular}}
\label{tab:0}
\end{table*}

\subsection{Pseudo Mask Generation and Noisy Label Training}
Based on the pseudo scribble, we now propose to create our trainable samples in unlabeled set $\{ \bm{X}_u, \bm{L}_u \}$ by generating pseudo masks $\bm{L}_u$ with GeoS \cite{criminisi2008geos, wang2018deepigeos} for each $[\bm{X}_u, \bm{Y}_u]$. $f_\theta$ is based on a 3D UNet \cite{ronneberger2015u} and we use the sum of cross entropy and dice loss \cite{milletari2016v} for supervision, which is commonly used in medical image segmentation. 
\par The experiments in Karimi \etal~\cite{karimi2020deep} suggest that iterative label cleaning method achieves the best performance for the pseudo mask training. According to its conclusion, we adopt the state-of-the-art label correction algorithm PLC (Progressive Label Correction) \cite{yi2021learning} that iteratively corrects labels and refines the model $f_\theta$ for training. For each point $\bm{c}_i$, if the prediction of $f_\theta$ is different from its pseudo mask $l(\bm{c}_i)$ and its confidence is above the threshold, $f_\theta(\bm{c}_i)>\delta$, the label $l(\bm{c}_i)$ is flipped to the prediction of $f_\theta$. We repeatedly correct masks and improve the network until it converged.
\par

\section{Experiments}


\subsection{Dataset and Evaluate Metric}
To demonstrate the generalization of our method, we perform evaluation under two highly different CT datasets: abdomen organs segmentation and HaN organs segmentation:
\par - The Cancer Image Archive (TCIA) Pancreas CT dataset \cite{clark2013cancer} contains 43 patients with various abdomen tumors. In practice, we test our method on 3 organs: liver, spleen and left liver\footnote{\href{https://zenodo.org/record/1169361\#.YXuabRrP2Uk}{https://zenodo.org/record/1169361\#.YXuabRrP2Uk}}. They make up about \textbf{8.5\%} of the total volume.
\par - Automatic Structure Segmentation for Radiotherapy Planning Challenge 2019\footnote{\href{https://structseg2019.grand-challenge.org/Home/}{https://structseg2019.grand-challenge.org/Home/}} (StructSeg19) task 1 contains 50 HaN CT scans from nasopharynx cancer patients. We test our method on 3 organs: brain stem, right/left parotid glands (right/left PG), which account for \textbf{0.1\%} voxels in total.
\par For both two datasets, we use 60\% of data for training, 20\% for validation, and the remaining 20\% are used for testing. We use the same evaluation metric Dice score as in previous works \cite{ouyang2020self, zhao2019data}. Dice score  (0-1, 0: mismatch; 1: perfectly match) measures the overlap of the prediction $\bm{M}$ and ground truth $\bm{G}$, and is defined as:
\begin{equation}
    Dice(\bm{M}, \bm{G})=\frac{2|\bm{M}\cap\bm{G}|}{|\bm{M}|+|\bm{G}|}
\end{equation}
\subsection{Implementation Details}
The whole framework is implemented with Pytorch \cite{paszke2019pytorch} and public available soon\footnote{\href{https://github.com/LWHYC/OneShot\_WeaklySeg}{https://github.com/LWHYC/OneShot\_WeaklySeg}}. Our model is trained with a NVIDIA GTX 1080 Ti GPU. The Adam optimizer~\cite{kingma2014adam} is used for training with batch size 8, initial learning rate $10^{-3}$ with a stepping decay rate of 0.9 per 10 epochs and 150 epochs totally. Along $zxy$ plane, TCIA and StructSeg19 images are resampled to $3\times1\times1$mm$^3$ and $1\times1\times1$mm$^3$, respectively. Fixed patch size $48\times128\times128$ voxels is applied for PRNet training in both datasets. We set $\tau=0.5 $ for DFD. $\delta$ is set as 0.95 for PLC at beginning and times 0.99 every epoch till it reaches 0.85. 
\par We focus on the task of segmentation using one volume labeled with scribbles. We select the first subject in training data as support volume and draw the scribbles manually, then apply PRNet propagating them on the remaining ones to generate pseudo scribbles and use GeoS to obtain the pseudo masks. We crop the target organs around the boundary of pseudo masks for training to reduce the class imbalance between foreground and background. We compare the performance of models trained on pseudo masks generated by original/noise-reduction pseudo scribbles in Sec. \ref{sec: 4.4}.
 \begin{figure}[t]
    \includegraphics[width=1\linewidth]{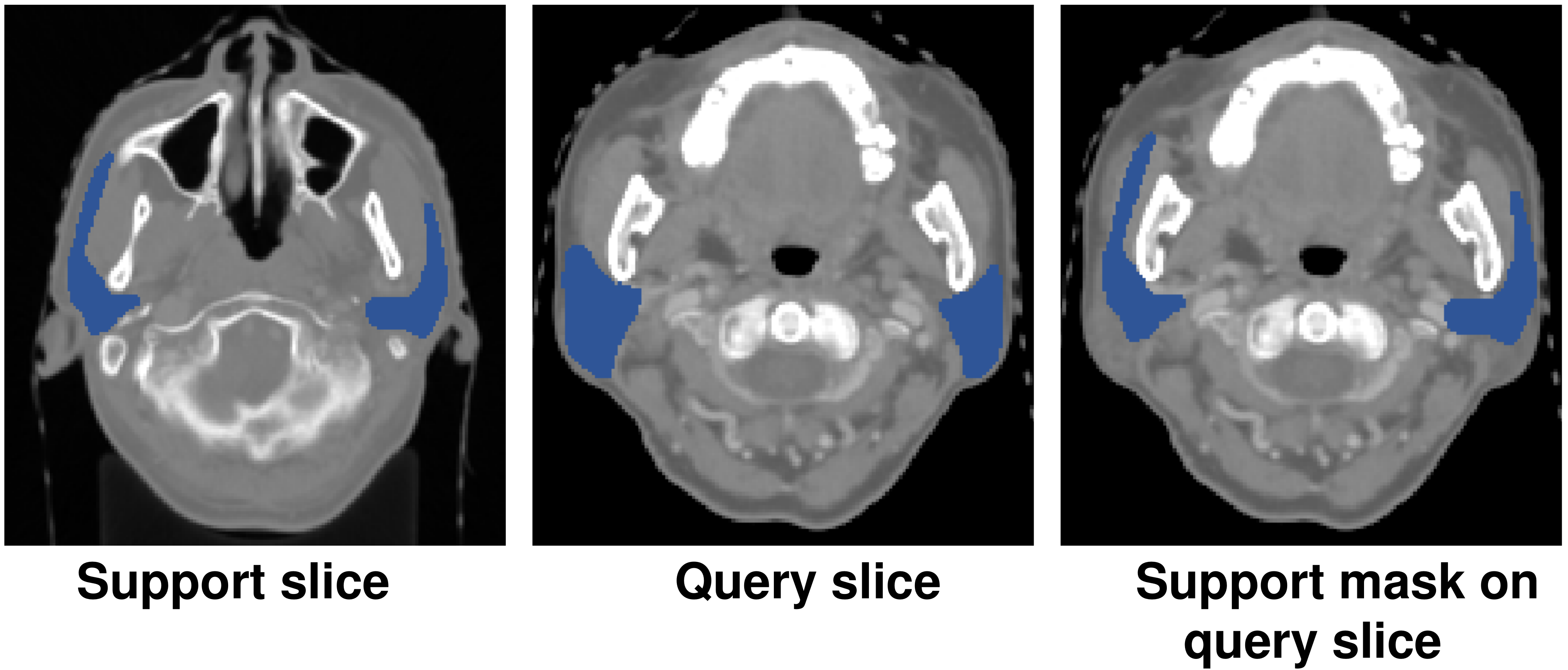}
   \caption{Visualization example of low intersection of parotid glands between support and query slices in StructSeg19. }
\label{fig:intersection}
\end{figure}
 \begin{figure}[t]
    \includegraphics[width=1\linewidth]{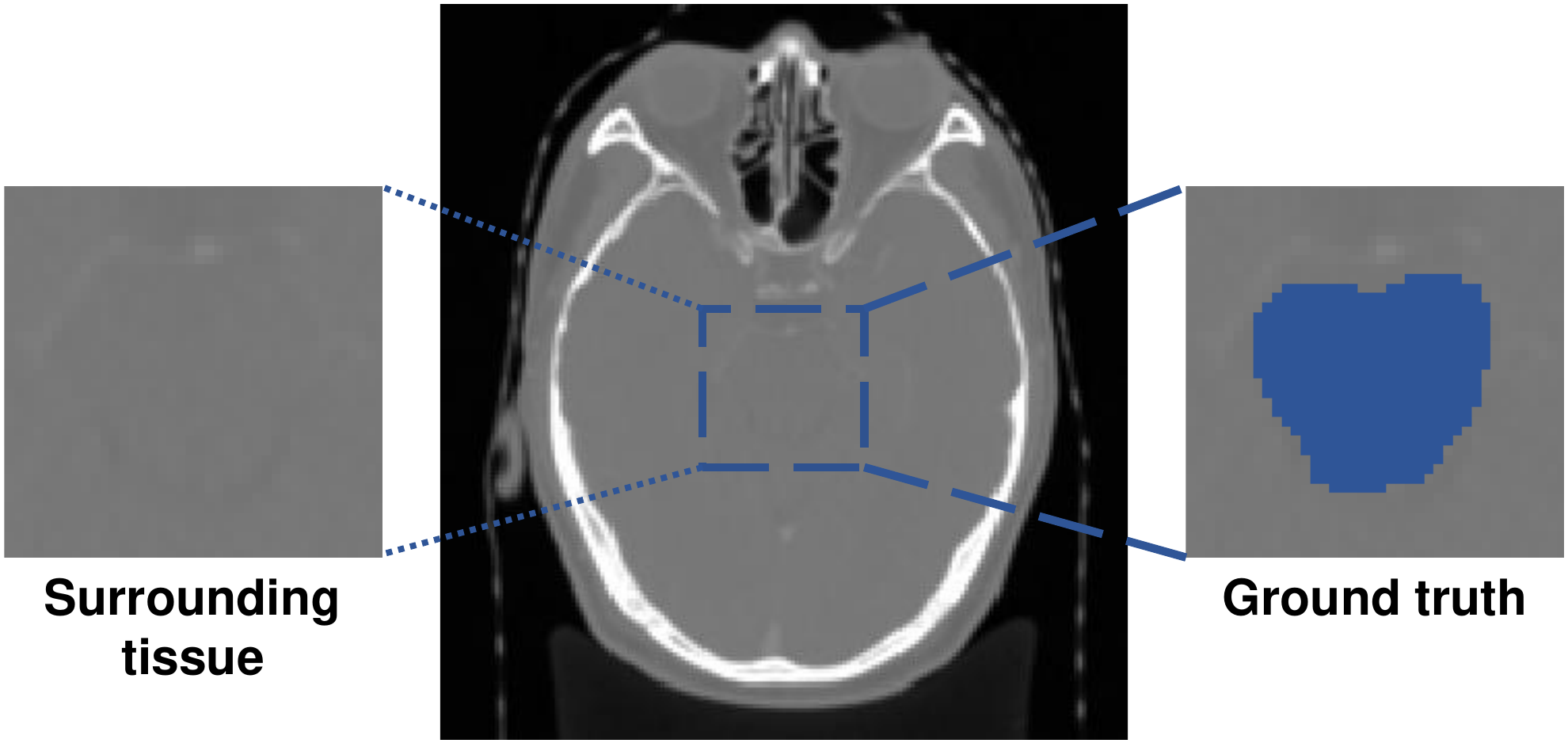}
   \caption{Visualization example of low contrast between brain stem and surrounding tissue in StructSeg19. }
\label{fig:contrast}
\end{figure}
 \begin{figure*}[ht]
    \includegraphics[width=1\linewidth]{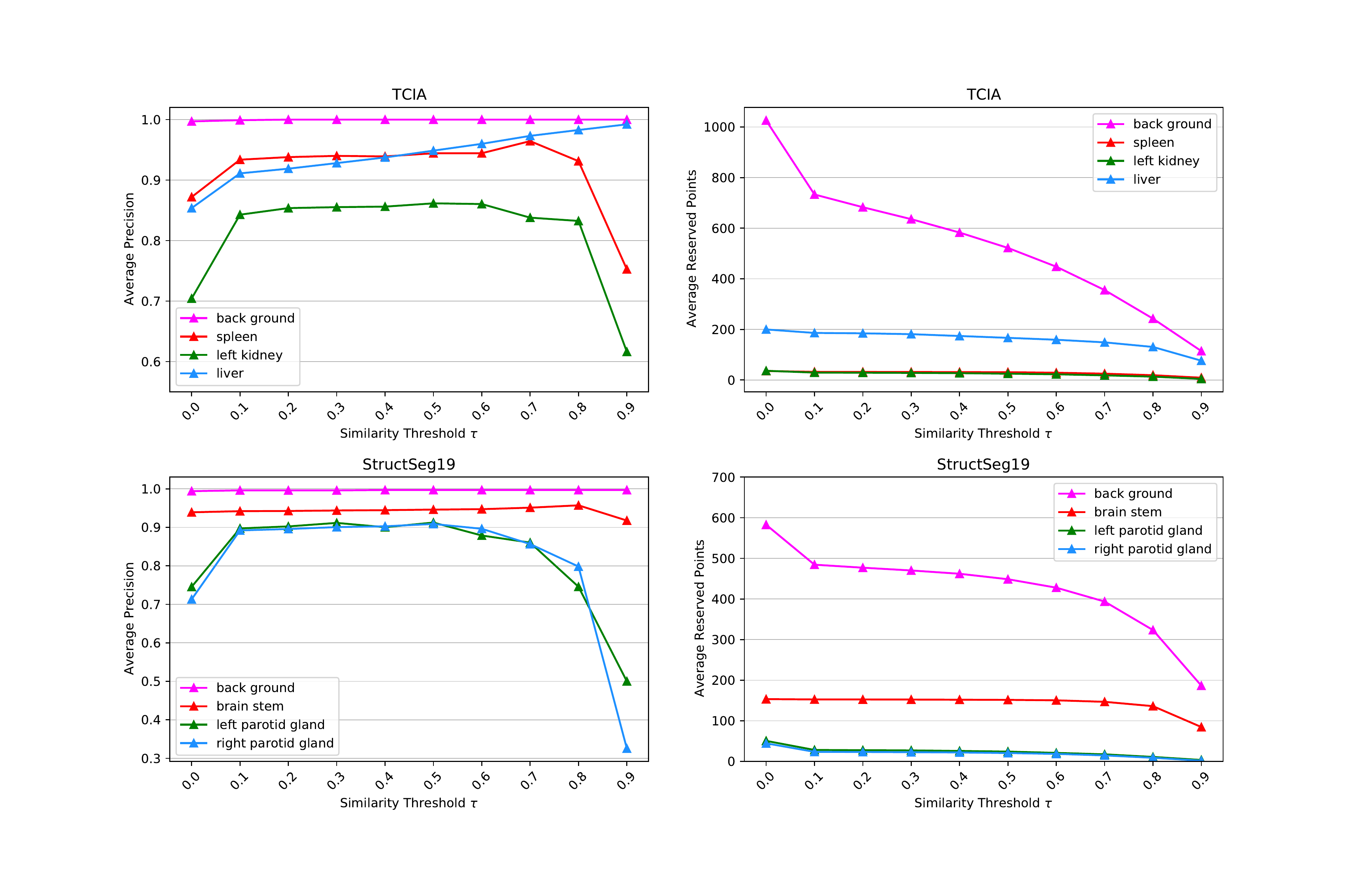}
   \caption{Average precision and the number of reserved points under different $\tau$ in validation set. }
\label{fig:sim_thresh}
\end{figure*}
\subsection{Comparison with State-of-the-art Methods}
Table \ref{tab:0} shows the comparisons of our method with 3 open-source OSL methods: SE-Net \cite{roy2020squeeze}, SSL-ALPNet \cite{ouyang2020self} and DataAug \cite{zhao2019data}, a data augmentation methods for synthesizing labeled medical image. For SE-Net and SSL-ALPNet, we adopt the same setting in their works, cropping the query volume among the bottom and top slice of the target organ first then dividing it into 3 equally-spaced chunks to be segmented with corresponding support slices. We also train a fully-supervised segmentation 3D UNet that uses ground truth labels for all examples in our training dataset to serve as the upper bound.

\par Without being indicated the range of slices where the organ lies, our proposed PRNet outperforms others largely, especially in StructSeg19. There are mainly 2 reasons: (1) SE-Net passes the support set through the conditioner arm, whose information is conveyed to the segmenter arm via interaction blocks. It assumes that the target organ in the support slice is roughly aligned with the one in the query. However, due to the limited volume size of organs in HaN, even slight intersecting pixels of target organs between support and query slices caused by small variance among patients could make the above assumption unsatisfied. Fig. \ref{fig:intersection} presents an visualization example of parotid glands. It could be observed that the propagation of the support mask on query slice has a small overlap with the ground truth; (2) SSL-ALPNet utilizes a self-supervised superpixel segmentation task then uses the learned representations to segment new classes without fine-tuning. However, as shown in Fig. \ref{fig:contrast}, the contrast of organs, e.g., brain stem in StructSeg19 with surrounding tissue, is extremely low even after normalization, which is vital to SSL-ALPNet because it could hardly select superpixels distinguishing the boundary of target organs during training time. In contrast, our method first locates fore-/background scribbles in query volumes, thus is not sensitive to the extreme sample imbalance and low contrast.


\begin{table*}[t!]
\centering
\caption{Ablation study (in Dice score) on $\tau$ values in testing set.}
\scalebox{0.9}{
\begin{tabular}{@{}ccccccccc@{}}
\toprule[1pt]
&\multicolumn{4}{c}{TCIA}& \multicolumn{4}{c}{StructSeg19} \\
$\tau$  & Spleen  & Left Kidney  & Liver & Mean & Brain Stem  & Left PG  & Right PG & Mean \\ \midrule
0.0 & 69.6$\pm$17.3 & 51.6$\pm$27.5 & 82.9$\pm$5.7 & 68.0$\pm$13.2 & \textbf{74.4$\pm$2.1} &53.3$\pm$24.0& 51.4$\pm$23.1 &59.7$\pm$14.6 \\
0.5 & \textbf{81.1$\pm$12.6} & \textbf{70.0$\pm$23.4}  & \textbf{83.2$\pm$5.5} & \textbf{78.1$\pm$11.6} & 74.0$\pm$2.0 & \textbf{64.7$\pm$11.3} & \textbf{65.9$\pm$15.0} & \textbf{68.2$\pm$7.7} \\
0.9  & 43.6$\pm$35.6 &21.9$\pm$25.1& 70.1$\pm$10.9 & 45.2$\pm$16.1 & 67.3$\pm$6.9 &9.2$\pm$23.8& 19.5$\pm$27.7 &32.0$\pm$16.1 \\
\bottomrule[1pt]
\end{tabular}}
\label{tab:3}
\end{table*}

\subsection{Ablation Study}
\textbf{Effect of DFD}
\label{sec: 4.4}
To verify the contribution of DFD, we conduct experiments at two stages: (1) scribble propagation; (2) pseudo masks generation with original/denoised scribbles.
\par
\begin{figure}[t!]
    \includegraphics[width=1\linewidth]{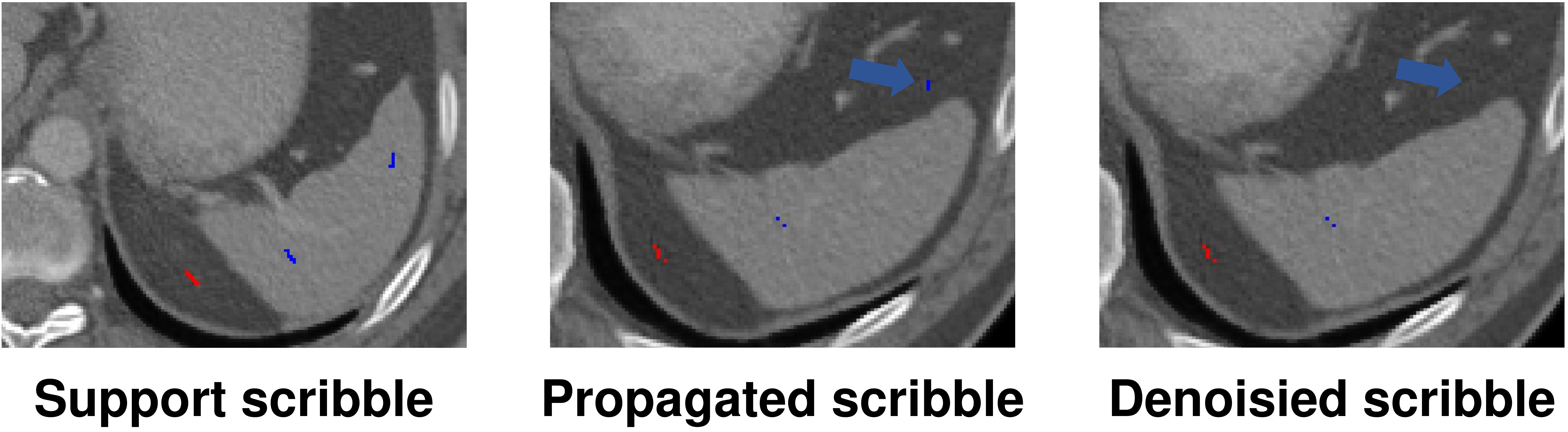}
   \caption{Visualization example of the denoised scribble. After denoising, spleen scribble propagation (blue) on the top right corner has no intersection with the background area, while the remaining correctly located spleen and background scribbles (red) are reserved.}
   \vspace{-1em}
\label{fig:denoising example}
\end{figure}

 First, to confirm the denoising ability of DFD and select the appropriate $\tau$, we gradually increase $\tau$ from 0 to 0.9 and evaluate the results in the validation set. Fig. \ref{fig:sim_thresh} shows the class-specific average precision and the number of reserved points per volume before and after denoising. For most classes, even be filtered with the lowest level ($\tau=0.1$), their propagation precision would increase largely. For example, left/right parotid glands with $\tau=0.1$ outperform the original results ($\tau=0$) by an average precision score around 20\%. A visual comparison example is presented in Fig. \ref{fig:denoising example}, in which DFD successfully deletes the incorrectly located spleen scribble (blue) while reserving others. 

\begin{table*}[t!]
\centering
\caption{Ablation study (in Dice score) on model training in testing set.}
\scalebox{0.87}{
\begin{tabular}{@{}lccccccccc@{}}
\toprule[1pt]
& &\multicolumn{4}{c}{TCIA}& \multicolumn{4}{c}{StructSeg19} \\
Method & Training & Spleen  & Left Kidney  & Liver & Mean & Brain Stem  & Left PG  & Right PG & Mean \\ \midrule
RDR+GeoS & \XSolidBrush & 81.1$\pm$12.6 & 70.0$\pm$23.4  & 83.2$\pm$5.5 & 78.1$\pm$11.6 & 74.0$\pm$2.0 & 64.7$\pm$11.3 & 65.9$\pm$15.0 & 68.2$\pm$7.7 \\
RDR+GeoS & \CheckmarkBold & \textbf{85.3$\pm$11.5} & 73.2$\pm$23.7  & 87.3$\pm$5.0 & 81.9$\pm$11.2 & 73.4$\pm$5.1 & 74.1$\pm$2.0 & \textbf{77.0$\pm$2.1} & 74.8$\pm$2.2 \\
RDR+GeoS+PLC &\CheckmarkBold & 84.9$\pm$12.0 & \textbf{90.9$\pm$4.1}  & \textbf{90.3$\pm$3.6} & \textbf{88.7$\pm$4.2} & \textbf{75.2$\pm$3.4} & \textbf{74.6$\pm$3.3} & 76.0$\pm$2.8 & \textbf{75.3$\pm$3.2} \\
\bottomrule[1pt]
\end{tabular}}
\vspace{-1em}
\label{tab:4}
\end{table*}
\par However, when $\tau$ reaches 0.9, it witnesses a steep decrease in both average precision and reserved points for almost all classes. The reason is that for some subjects, $\tau$ with such a high value may remove all propagated points, leading to their precision and the number of reserved points equal to 0. Therefore, to balance the precision and the number of reserved points, we set $\tau=0.5$.

\par Another interesting phenomenon is that for both datasets, the precision of background scribbles remains approximately equal to 1. It's mainly because of the extreme imbalance between background and foreground pixels. The background part takes up nearly 91.5\% in TCIA and 99.9\% in StructSeg19 thus the propagated points are highly likely to fall into the background area. 
\par Second, we explore how the noise reduction affects the accuracy of pseudo masks. We compare the pseudo masks generated by GeoS from scribbles filtered with different $\tau=0, 0.5, 0.9$.
As shown in Table \ref{tab:3}, the $\tau=0.5$ in subsequent experiments brings around 10\% improvement than original results in both datasets. As expected in Sec. \ref{sec 3.3}, improving the precision of located points rather than the total number brings higher pseudo masks accuracy. The results in Table \ref{tab:3} show that our method could yield competitive performance even without further training.

\par
\textbf{Effect of Noisy Training}
 \begin{figure}[t!]
    \includegraphics[width=1\linewidth]{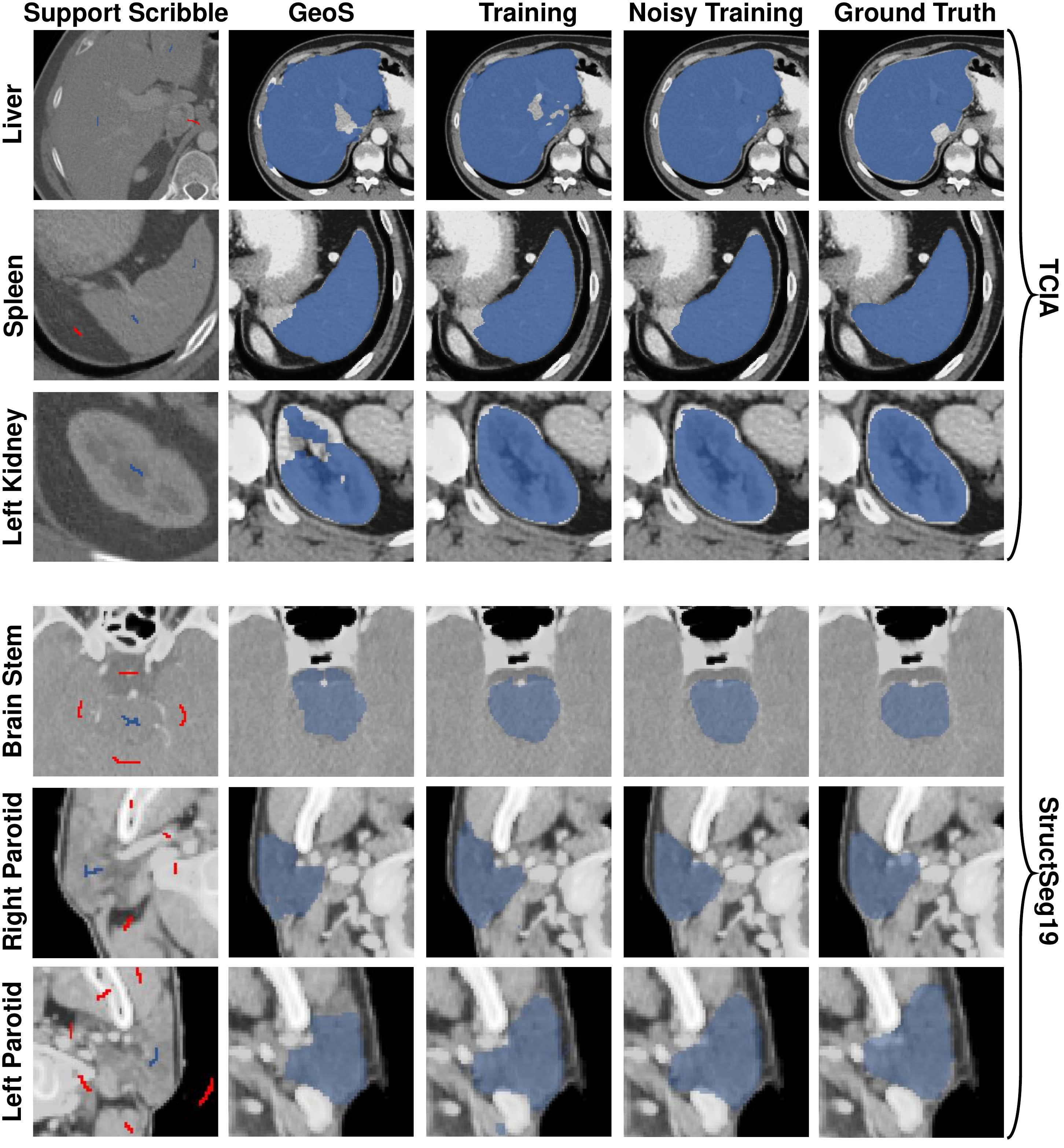}
   \caption{Examples of predictions under different stages.}
   \vspace{-1em}
\label{fig:visualization}
\end{figure}
To show the necessity and advantages of the noisy training, we conduct experiments training $f_\theta$ with and without PLC, shown in Table \ref{tab:4}. From the table, we can observe that $f_\theta$ brings a 3.8\% and 6.6\% improvement in TCIA and StructSeg19 compared to the pseudo masks. This implies that even without any specific setting, deep CNNs are robust to strong label noise. 
With iterative label cleaning, PLC brings an extra 6.8\% improvement in TCIA and 0.5\% in StructSeg19.
\par Fig. \ref{fig:visualization} shows segmentation examples under different settings. It could be noted that even with very sparse support annotations, our training model could yield precise results.
\section{Conclusion and Discussion}
In this work, we present a novel one-shot medical image segmentation framework, which incorporates one-shot localization and weakly-supervised segmentation. Given one image labeled with scribbles and plenty of unlabeled volumes, the proposed method first locates support scribbles on each image, then filters them with proposed dual-level feature denoising and applies WSS algorithms to build a large training set. We use the generated masks to train a supervised segmentation model with noisy training algorithms for every class. Experiments on two public datasets demonstrate that without manual localization, our method outperforms existing OSS models largely and could perform robustly even under the extreme sample imbalance. What's more, our method could be easily promoted with future improvement in WSS algorithms and noisy training algorithms. And the proposed DFD could be used for judging the correctness of the results in any landmark or organ localization tasks.
\section{Limitations}
Since the proposed PRNet is based on the assumption that different people share similar anatomical structures, it may not yield satisfying results under extreme situations. For example, our model trained on the scans of adults may fail in locating and segmenting organs on scans of infants. But it could be resolved by fine-tuning the model with several unlabeled target scans.

{\small
\bibliographystyle{ieee_fullname}
\bibliography{egbib}
}
\end{document}